\documentclass[12pt,nohyper]{JHEP}          
\usepackage{epsfig}
\usepackage{amssymb}
\newcommand{\be}{\begin{equation}}
\newcommand{\ee}{\end{equation}}
\newcommand{\bea}{\begin{eqnarray}}
\newcommand{\eea}{\end{eqnarray}}
\newcommand{\bref}[1]{(\ref{#1})}
\newcommand{\nn}{\nonumber}
\newcommand{\A}{\alpha} \newcommand{\B}{\beta} 
 \newcommand{\D}{\delta} 
\newcommand{\ep}{\epsilon} \newcommand{\vep}{\varepsilon}
\newcommand{\T}{\theta} 
   \newcommand{\vp}{\varphi}
         \newcommand{\lam}{\lambda}
\renewcommand{\r}{\rho}           \newcommand{\s}{\sigma}

\newcommand{\h}{\eta}

         \newcommand{\Lam}{\Lambda}
\newcommand{\iPi}{{\it\Pi}}

\newcommand{\ba}{\overline }
\def\6{\partial}
\def\7{\tilde}
\def\8{\widehat}
 \def\bx{{\bf x}}

\def\pa{\partial} 
 
\def\dpo{d {\hskip -0.02cm{+}} {\hskip -0.05cm{1}}}

\def\CC{{\cal C}}\def\CP{{\cal P}}\def\CD{{\cal D}}\def\CL{{\cal L}}
\def\CA{{\cal A}}\def\CF{{\cal F}}\def\CQ{{\cal Q}}\def\CE{{\cal E}}
\def\CB{{\cal B}}\def\cf{{\cal K}}
\def\l{{\ell}}

\def\lag{Lagrangian }\def\lags{Lagrangians }
\def\ham{Hamiltonian }
\def\NC{{\it NC} }

\title{Gauge and BRST Generators for Space-Time Non-commutative U(1) Theory
}
\author{Joaquim Gomis 
\\Theory Division, CERN, CH-1211 Geneva 23, Switzerland
\\Departament ECM, Facultat de F\'{\i}sica
\\Universitat de Barcelona and Institut de F\'{\i}sica d'Altes Energies,
\\Diagonal 647, E-08028 Barcelona, Spain
\\Email: {\tt gomis@ecm.ub.es }}
\author{Kiyoshi Kamimura 
\\Department of Physics, Toho University, 
\\Funabashi\ 274-8510, Japan \\
Email: {\tt kamimura@ph.sci.toho-u.ac.jp}}
\author{Toni Mateos 
\\Departament ECM, Facultat de F\'{\i}sica
\\Universitat de Barcelona and Institut de F\'{\i}sica d'Altes
Energies,
\\Diagonal 647, E-08028 Barcelona, Spain
\\Email: {\tt tonim@ecm.ub.es}}

\abstract{ The \ham (gauge) symmetry generators of non-local (gauge) theories 
are presented.
The construction is based on the $d+1$ dimensional space-time 
formulation of $d$ dimensional non-local theories. 
The procedure is applied to $U(1)$ space-time non-commutative gauge 
theory. In the \ham formalism the \ham and the gauge generator are 
constructed. The nilpotent BRST charge is also obtained. 
The Seiberg-Witten map between non-commutative and commutative theories 
is described by a canonical transformation in the
superphase space and in the field-antifield space. 
The solutions of classical master equations for non-commutative and 
commutative theories are related by a canonical transformation in the 
antibracket sense.
} 

\preprint{CERN-TH/2000-246, UB-ECM-PF-00/10, TOHO-FP-0068}

\begin{document}
\section{Introduction}
\indent

Non-local theories are described by actions that contain an infinite 
number of temporal derivatives. There exists an equivalent formulation 
of those theories in a space-time of one dimension higher \cite{lv}. 
In this formulation there are two time coordinates,  
and the dynamics in this space is described
in such a way that the evolution is local with respect to one of the times. 
Thanks to this, a  \ham formalism can be constructed in the 
$\dpo$ dimensions as a local 
theory with respect to the evolution time
\cite{lv}\cite{gkl}\cite{woodd1}\cite{bearing}, in which the Euler-Lagrange 
equations appear as \ham constraints \cite{gkl}.
A characteristic feature is that there is no dynamics in the usual sense; 
{\it i.e.} the physical trajectories are not obtained as evolution of 
some given initial conditions. 

In this paper we construct the symmetry generators for non-local theories. 
Corresponding to symmetries of a non-local \lag the symmetry generators 
are constructed in a natural way in $\dpo$ dimensions and are conserved 
quantities. When the original symmetries of the non-local theory are gauge 
symmetries the corresponding transformations are realized as {rigid} 
symmetries in the $\dpo$ dimensions.   

We analyze in detail the case of space-time non-commutative (\NC)
$U(1)$ gauge theory\footnote{Here we use the term "$U(1)$" for 
"rank one" gauge field. It is not abelian for the \NC case.}. 
In particular, we obtain its Hamiltonian and we show that
it is the generator of time translations.
We also study the relation between the gauge generators of the \NC and 
commutative theories, by considering the Seiberg-Witten (SW) map \cite{sw}
as an ordinary canonical transformation.

We then move to study the BRST symmetry of this $U(1)$ $NC$ theory
and we construct the nilpotent \ham BRST charges .
We also analyze the BRST symmetry 
at Lagrangian level using the field-antifield formalism.
The Seiberg-Witten (SW) map \cite{sw} is extended 
to a canonical transformation in superphase space and 
in the field-antifield space. We show that the 
solutions of the classical master  equation 
for non-commutative and commutative theories are related by 
a canonical transformation in the antibracket sense.

The organization of the paper is as follows. In section 2 we study the
general properties of symmetry generators of non-local theories. In
section 3 we construct the gauge symmetry generator for $U(1)$
\NC gauge theory. Section 4 is devoted to study the relation between 
the gauge generators of commutative and $U(1)$ \NC gauge theories. 
In section 5 we construct the BRST generator. There 
is an appendix where the ordinary $U(1)$ local Maxwell theory is analyzed 
in terms of the $\dpo$ dimensional formalism. 

\section{Hamiltonian formalism of non-local theories and 
symmetry generators}
\indent

\subsection{Brief Review}

A non-local \lag  at time $t$ depends not only on variables at time $t$ 
but also on ones at different times. In other words it depends on an 
infinite number of time derivatives of the positions $q_i(t)$ \footnote
{For simplicity, in this section we will explicitly consider the case of mechanics.}.
The analogue of the tangent bundle for \lags depending on positions 
and velocities is now infinite dimensional. It is the space of all 
possible trajectories. 
The action is
\be
S[q]=\int dt ~L^{non}(t).
\label{action}
\ee
The Euler-Lagrange 
(EL) equation is obtained by taking the functional variation of 
\bref{action},
\be
\frac {\delta S[q]}{\delta q{_i}(t)}~=~\int dt' E{^i}(t',t;[q])
~=~0,~~~~~~~~ 
E{^i}(t',t;[q])~\equiv~\frac {\delta L^{non}(t')}{\delta q{_i}(t)}.
\label{EL}
\ee

One of the most striking features of such theories is that of the new
interpretation that this EL equation has
\cite{lv}\cite{gkl}. Since the equations
of motion are of infinite degree in time derivatives, one should give as
initial conditions the value of all these derivatives at some initial time.
In other words, we should give the whole trajectory (or part of it) 
as the initial condition. If we denote the space of all possible
trajectories as $J=\{q(\lambda),\lambda \in R \}$, then \bref{EL} is
a Lagrangian constraint defining the subspace $J_R \subset J$ of
physical trajectories.

In \cite{lv}\cite{gkl} this was
implemented using a formalism in which one works with one extra dimension.
The final result was that one obtains a two dimensional field theory 
whose Lagrangian is 
\bea
\tilde{L}(t,[\CQ]) := \int \; d\lambda \; \delta(\lambda) \; \CL(t,\lambda)
\eea
where the Lagrangian density $\CL(t,\lambda)$ is constructed from the 
original non-local one $L^{non}$ by performing the following replacements
\bea
q_i(t)&\to&\CQ_i(t,\s),~~~ 
\frac{d^n}{dt^n}q_i(t)~\to~\frac{\pa^n}{\pa\s^n}\CQ_i(t,\s),~~~~
q_i(t+\rho)~\to~\CQ_i(t,\s+\rho).
\label{rep}
\eea
Note that this $1+1$ field theory has two ``time'' coordinates $t$ and
$\sigma$ but, using these replacements, the dynamics is described in such
a way that the evolution is local with respect to one of them ($t$).
This is the key achievement that will enable us to analyze
many aspects of the $1+1$ theory using ordinary methods from
local theories. 

The theory was also shown to have the following Hamiltonian
\bea
H(t)&=&\int d\s ~[~\CP{^i}(t,\s)\CQ{_i}'(t,\s)~-~\D(\s)\CL(t,\s)~],
\label{h}
\eea
where $\CQ_i'(t,\s)\equiv\pa_\s \CQ_i(t,\s)$ and $\CP{^i}(t,\s)$ are the
canonical momenta. Note that the \ham
depends on the fields $\CQ_i(t,\s)$ 
and an infinite number of sigma derivatives of it, 
but not on any derivative with respect to the evolution time $t$.
Thus the \ham \bref{h} is indeed a well defined phase space quantity.

The Hamilton equations are, denoting time ($t$) derivatives by "dots",
\bea
\dot \CQ{_i}(t,\s)&=& \CQ_i'(t,\s),
\label{Qdot}\\
\dot \CP{^i}(t,\s)&=& {\CP^i}'(t,\s)+\frac{\D \CL(t,0)}{\D \CQ{_i}(t,\s)}=
{\CP^i}'(t,\s)+\CE{^i}(t;0,\s),
\label{Pdot}
\eea
where $\CE(t;\s',\s)$ is defined by
\bea
\CE{^i}(t;\s',\s)&=&\frac{\D \CL(t,\s')}{\D \CQ{_i}(t,\s)}.
\eea
Equations \bref{Qdot} 
restrict the two dimensional fields $ \CQ_i(t,\s)$ to depend
only on a chiral combination of the two times  $t+\s$ on shell.
They are 
identified with the position variables $q_i(t)$ of the original system by
\bea
\CQ_i(t,\s)&=& q_i(t+\s),~~~~~~i.e.~~~~~~~q_i(\sigma)~=~\CQ_i(0,\sigma).
\label{qdef}
\eea

The solutions to these $1+1$ dimensional
field equations are related to those of the EL equations \bref{EL}
of the original non-local Lagrangian $L^{non}$ if we impose
a constraint on the momentum \cite{lv}
\bea
\vp{^i}(t,\s)&=&\CP{^i}(t,\s)~-~\int d\s'~\chi(\s,-\s')~\CE{^i}(t;\s',\s)~
\approx~0,
\label{vp}
\eea
where $ \chi(\s,-\s')$ is defined by using the sign distribution $\ep(\s)$
as $\chi(\s,-\s')~=~\frac{\ep(\s)-\ep(\s')}{2}$.
We use {\it weak equality} symbol "$\approx$" for equations
those hold on the constraint surface
\cite{Dirac}. As usual, one has to impose stability to this constraint,
leading us to the following one
\bea
\dot\vp{^i}(t,\s)~\approx~\D(\s)~[\int d\s'~\CE{^i}(t;\s',0)]\approx 0.
\label{EOM}
\eea
We should require  further consistency conditions of this constraint. 
Repeating this we get an infinite set of Hamiltonian constraints 
which can be expressed collectively as
\bea
\tilde\vp{^i}(t,\s)=\int d\s'\CE{^i}(t;\s',\s)&\approx&0
,~~(-\infty<\s<\infty).
\label{EOMf}
\eea
If we use \bref{Qdot} and \bref{qdef} it reduces to the EL equation 
\bref{EL} of $q_i(t)$ obtained from $L^{non}(t)$.
This is precisely what we were seeking at the beginning, since now
we see that the new 1+1 Hamiltonian system incorporates the EL equation
as a constraint on the phase space.

	Summarizing, we have been able to describe the original
non-local \lag system as a $1+1$ dimensional {local} (in one 
of the times) \ham  system, governed by the   \ham \bref{h} 
and  the constraints \bref{vp} 
and \bref{EOMf}. The formalism introduced here can be thought of as
a generalization of the Ostrogradski 
formalism \cite{o} to the case of infinite derivative theories.
\medskip
\subsection{Hamiltonian symmetry generators}
\indent

For local theories symmetry properties of the system are examined 
using the N\"oether theorem \cite{nother}.  
In \ham formalism the relation between
symmetries and conservation laws has been discussed extensively for 
singular Lagrangian systems, for example
\cite{kp}\cite{GKP}\cite{josepmaria}. 
In this section, we develop a formalism to treat the case of non-local theories.

Suppose we have a non-local \lag, \bref{action}, which  is invariant under 
some transformation $\D q(t)$ up to a total derivative,
\bea
\D L^{non}(t)&=&\int dt'~\frac{\D L^{non}(t)}{\D q{_i}(t')}~
{\D q{_i}(t')}~=~\frac{d}{dt}k(t).
\label{DL}
\eea
Now we move to our $1+1$ dimensional theory and take profit
of the fact that it was local in the evolution time $t$. 
Therefore, we can construct the corresponding symmetry 
generator in the \ham formalism in the usual way
\bea
G(t)&=&\int d\s ~[~\CP{^i}(t,\s)\D \CQ{_i}(t,\s)~-~\D(\s)\cf(t,\s)~],
\label{G}
\eea
where $\D \CQ_i(t,\s)$ and  $\cf(t,\s)$ are constructed from
 $\D q(t)$ and  $k(t)$ respectively by the same replacement \bref{rep},
as $\CL(t,\s)$ was obtained from $L^{non}(t)$. The quasi-invariance of the
non-local \lag \bref{DL}, translated to the $1+1$ language,  means  
\bea
 \int d\s'~\frac{\D\CL(t,\s)}{\D \CQ{_i}(t,\s')}{\D \CQ{_i}(t,\s')}&=&
\pa_\s \cf(t,\s).
\label{paF}
\eea

When the original non-local \lag has a gauge symmetry the $\D q_i(t)$ 
and  $k(t)$ contain an arbitrary function of time $\lam(t)$ and its 
$t$ derivatives. 
In $\D \CQ_i(t,\s)$ and  $\cf(t,\s)$ the $\lam(t)$ is replaced  by 
$\Lam(t,\s)$ in the same manner  as  $q_i(t)$ is replaced  by 
$\CQ_i(t,\s)$ in \bref{rep}. 
However in order for the transformation generated by \bref{G} to be 
a symmetry of the Hamilton equations, 
$\Lam(t,\s)$ can not be an arbitrary function of $t$ but should satisfy
\bea
\dot\Lam(t,\s)&=&\Lam'(t,\s)
\label{Lamdd}
\eea
as will be shown shortly.
This restriction on the parameter function $\Lam$ means that the
transformations generated by $G(t)$ in the $\dpo$ dimensional
\ham formalism are {rigid} transformations in contrast with the 
original ones for the non-local theory which are {gauge} transformations.
In the appendix we will see how this {rigid} transformations in
the $\dpo$  dimensional \ham formalism are reduced to the usual 
{gauge} transformations
in $d$ dimension for the $U(1)$ Maxwell theory.
\medskip

The generator $G(t)$ generates the transformation of $\CQ_i(t,\s)$,
\bea
\D \CQ{_i}(t,\s)&=&\{\CQ{_i}(t,\s),~G(t)\},
\eea
corresponding to the transformation $\D q_i(t)$ in the non-local Lagrangian. 
It also generates the transformation of the 
momentum $\CP^i(t,\s)$ and so, of any functional of the phase space variables.
In particular, we will see that, as consistency demands, the Hamiltonian 
\bref{h}
and the constraints \bref{vp} and \bref{EOMf} are 
invariant, in the sense
that their symmetry 
transformation vanishes on the hypersurface of
phase space determined by the constraints.
Let us state a series of results and properties of our gauge generator.

\medskip

{\it a) G(t) is a conserved quantity}
\bea
\frac{d}{dt}G(t)&=&\{G(t),H(t)\}~+~\frac{\pa}{\pa t}G(t)
\label{dtG1}\\
&=& \int d\s d\s'\left[~\CP{^j}(t,\s) \left( \frac{\D(\D \CQ{_j}(t,\s))}
{\D \CQ{_i}(t,\s')}\CQ{_j}'(t,\s')-
\pa_\s\D(\s-\s'){\D \CQ{_j}(t,\s')}
\right.\right.\nn\\ &&\left.\left.~~~~~~~~~+~
\frac{\D(\D \CQ{_j}(t,\s))}{\D \Lam(t,\s')}\dot\Lam(t,\s')\right)~-~
\D(t,\s)~\left(\frac{\D \cf(t,\s)}{\D \CQ{_i}(t,\s')}\CQ{_i}'(t,\s')
\right.\right.\nn\\ &&\left.\left.~~~~~~~~~-~
\frac{\D(\CL(t,\s))}{\D \CQ{_i}(t,\s')}{\D \CQ{_i}(t,\s')}+
\frac{\D \cf(t,\s)}{\D \Lam(t,\s')}\dot\Lam(t,\s')\right)\right]~=~0.
\label{dtG}\eea
The last term of \bref{dtG1} is an explicit $t$ derivative 
through $\Lambda(t,\s)$. In order to show \bref{dtG} 
we need to use the symmetry condition \bref{paF} and 
the condition on $\Lambda(t,\s)$ in \bref{Lamdd}.

\medskip
\medskip

{\it b) All the constraints are invariant under the symmetry transformations.}

\medskip

	Let us show first the invariance of \bref{EOMf}, which is
nothing but the invariance of the equations of motion, as was to expected
for $G(t)$ generating a symmetry,

\bea
\{\7\vp{^i}(t,\s),G(t)\}&=&\{\int d\s''\CE{^i}(t,\s'',\s),
\int d\s' ~[~\CP{^j}(t,\s')\D \CQ{_j}(t,\s')~-~\D(\s')\cf(t,\s')~]\}
\nn\\
&=&\int d\s' d\s'' \frac{\D^2\CL(t,\s'')}{\D \CQ{_j}(t,\s')\D \CQ{_i}(t,\s)}
\D \CQ{_j}(t,\s')~=~
\int d\s' \frac{\D \7\vp{^j}(t,\s')}{\D \CQ{_i}(t,\s)}\D \CQ{_j}(t,\s')
\nn\\
&=&-~\int d\s'  \7\vp{^j}(t,\s')\frac{\D (\D \CQ{_j}(t,\s'))}
{\D \CQ{_i}(t,\s)}~
\approx~0,
\label{vptG}
\eea
where we have used an identity obtained from \bref{paF}, 
\bea
\int d\s d\s'\CE{^j}(t,\s,\s')\D \CQ{_j}(t,\s')&=&
\int d\s'\7\vp{^j}(t,\s')\D \CQ{_j}(t,\s')~=~0.
\label{IpaF}
\eea

Let us show now the invariance of the other constraint 
\bref{vp}. Using \bref{paF} and \bref{IpaF},
\bea
&&\{\vp{^i}(t,\s),G(t)\}=
\nn\\&&
=-\int d\s'\vp{^j}(t,\s')\frac{\D (\D \CQ{_j}(t,\s'))}{\D \CQ{_i}(t,\s)}-
\int d\s'[\int d\s''\chi(\s',-\s'')\CE{^j}(t;\s'',\s') 
\frac{\D (\D \CQ{_j}(t,\s'))}{\D \CQ{_i}(t,\s)}
\nn\\&&~\hskip 22mm~
-\D(\s')
\frac{\D(\cf(t,\s'))}{\D \CQ{_i}(t,\s)}+
\int d\s''\chi(\s,-\s'') \frac{\D\CE{^i}(t;\s'',\s)}{\D \CQ{_j}(t,\s')}
\D \CQ{_j}(t,\s')]
\nn\\
&&
=-\int d\s'\vp{^j}(t,\s')\frac{\D (\D \CQ{_j}(t,\s'))}{\D \CQ{_i}(t,\s)}+
\int d\s'\chi(\s,-\s')\7\vp{^j}(t,\s') 
\frac{\D (\D \CQ_{j}(t,\s'))}{\D \CQ{_i}(t,\s)}~\approx~0.
\label{vpG}
\eea
Thus we have shown that the constraint surface defined by 
$\vp\approx\7\vp\approx 0$ 
is invariant under the transformations generated by $G(t)$.

\medskip
\medskip

{\it c) Our Hamiltonian \bref{h} is the generator of time translations.}

\medskip

Consider a non-local Lagrangian in \bref{action} that does not depend on 
$t$ explicitly, so that time translation is a symmetry of the Lagrangian.  
To show that the generator of such a symmetry is our Hamiltonian $H$ in 
\bref{h} and that it is conserved,
we should simply show that we recover its expression \bref{h} from the general 
form of the generator \bref{G}. Indeed,
the \lag changes as $\D L^{non}=\vep \dot L^{non}$ under a time translation
$\D q_i(t)=\vep \dot q_i(t)$. The corresponding generator in the present 
formalism is, using \bref{G}
\bea
G_H(t)&=&\int d\s ~[~\CP{^i}(t,\s)(\vep  \CQ{_i}'(t,\s))~-~\D(\s)(\vep\CL 
(t,\s))~],
\label{GH}
\eea
which is $\vep$ times the Hamiltonian \bref{h}.
In this case the conservation of the constraints \bref{vp} and \bref{EOMf}  
is understood also from \bref{vpG} and \bref{vptG}.
Our Hamiltonian in the $1+1$ theory being the generator of time
translations is telling us that we should consider it as giving
the energy of the system. Actually, as we show in the appendix for the $U(1)$
commutative case, if we were working in this $d+1$ formalism 
but for a {\it local} theory, we can always use the system of constraints
to reduce the redundant extra coordinates and obtain
the ordinary \ham of the local theory in $d$ dimensions. 
Nevertheless, for a truly non local theory, there is no such a
simplification and the phase space is infinite dimensional.
Our discussion then shows that
it is the Hamiltonian \bref{h} that we should
use for computing the energy of the system.

\medskip

To summarize this chapter, we have constructed the \ham symmetry generators of
a general non-local theory working in a $\dpo$ dimensional space.  In
this formulation original {gauge} symmetries in $d$ dimensions
are {rigid} symmetries in the $\dpo$ dimensional space. This way of
understanding of
gauge symmetries is also useful for ordinary higher derivative
theories, see appendix and \cite{gkll2}.
The rest of this paper will be mainly devoted to illustrate how our
formalism is applied to the case of the non commutative $U(1)$ theory.

\section{ $U(1)$ non-commutative gauge theory} 
\indent

\subsection{Brief review}

The  magnetic $U(1)$ non-commutative (NC) 
gauge theory appears in the decoupling limit of
D-p branes in the presence of a constant NS-NS two form \cite{sw}. The
theory
could formally be extended to the electric case. However in this case
the field theory is acausal \cite{sst}\cite{ab} and 
non-unitary \cite{gm}\cite{agm}. In  terms of strings this is 
because  there is an obstruction to the decoupling limit in the case
of an electromagnetic background 
\cite{sst1}\cite{gms}\cite{br}\cite{gmms}\cite{km}. 
Here we are interested in the most
general case of {\it space-time} non-commutativity with
$\theta^{0i}\neq 0$. {\footnote {A \ham formalism for the magnetic
theory ($\theta^{0i}=0$) is analyzed in \cite{Dayi}.}}

We consider the $U(1)$ (rank one) \NC  Maxwell theory
in $d$ dimensions  with the action 
\bea
S&=&\int d^dx~(-\frac14~\8 F_{\mu\nu}\8 F^{\mu\nu}),
\label{Maxwell}
\eea
where $\8 F_{\mu\nu}$ is the field strength of the  $U(1)$ \NC
gauge potential $\8 A_\mu$ defined by{\footnote
{We put "hats" on the quantities of the \NC theory.}}
\bea
\8 F_{\mu\nu}&=&\pa_\mu \8 A_\nu-\pa_\nu \8 A_\mu-
i[\8 A_\mu,\8 A_\nu].
\label{defF}
\eea
The commutators in this paper are defined by the Moyal * product as
\bea
[f,g]&\equiv&f*g-g*f,~~~~~~
f(x)*g(x)=[e^{i\frac{\T^{\mu\nu}}{2}\pa_{\A^\mu}\pa_{\B^\nu} }
f(x+\A)g(x+\B)]_{\A=\B=0}.
\label{Moyal}
\eea
The EL equation of motion is 
\bea
\8 D_\mu\8 F^{\mu\nu}&=&0,
\label{ELeq}
\eea
where the covariant derivative is defined by
$\8 D=\pa-i[\8 A,~~]$.
\medskip

The gauge transformation is 
\bea 
\D \8 A_\mu&=&\8 D_\mu \8\lam \label{u1trans}
\eea
and it satisfies a non-Abelian gauge algebra, 
\bea
(\D_{\8\lam} \D_{\8\lam'} -\D_{\8\lam'}\D_{\8\lam}) \8 A_\mu&=&
-i\8 D_\mu [\8\lam,\8\lam'].
\eea
Since the field strength transforms covariantly as
\bea
\D \8 F_{\mu\nu}&=&-i[ \8 F_{\mu\nu},\8\lam]
\eea
the \lag density of \bref{Maxwell} transforms as
\bea
\D (-\frac14~\8 F_{\mu\nu}\8 F^{\mu\nu})&=& 
\frac{i}{2}~[\8 F_{\mu\nu},\8\lam]~\8 F^{\mu\nu}.
\label{Dlag}
\eea 
Using $\int d x(f*g)=\int d x(fg)$ and the associativity of the 
{\it star} product \bref{Dlag}  becomes a total divergence, as
was to be expected for \bref{u1trans} being a symmetry. So the action 
\bref{Maxwell} is invariant under the $U(1)$ \NC transformations.
\medskip

\subsection{Going to the $d+1$ formalism}

The \lag \bref{Maxwell} contains time derivatives 
of infinite order and  is  non-local. 
The \NC gauge transformation \bref{u1trans} is also non-local since,
for electric backgrounds ($\theta^{0i}\neq 0$), 
it contains time derivatives of infinite order in $\lam$.
Let us now proceed to construct the Hamiltonian and the generator
for the $U(1)$ \NC theory using the
formalism introduced in the last section.
The canonical structure will be realized in the $\dpo$ dimensional formalism.
Corresponding to the $d$ dimensional gauge potential $\8A_\mu(t,\bx)$,
we denote the gauge potential in $\dpo$ dimensional one as 
$\8\CA_\mu(t,\s,\bx)$.
\footnote{From now on we will use calligraphic letters for
fields in the $d+1$ formalism.}
We regard $t$ as the evolution ``time''. Now 
$x^0=\s$ is the coordinate denoted by $\s$
of $q_i(t,\s)$ in the last section. 
The other $(d-1)$ spatial coordinates ${\bf x}$ correspond to
the indices $i$ of $q_i(t,\s)$. 
The signature of $\dpo$ space is $(-,-,+,+,...,+)$.

The canonical system equivalent to the non-local action \bref{Maxwell}
is defined by the Hamiltonian \bref{h} and two constraints,
 \bref{vp} and \bref{EOMf}. For our present theory, 
the Hamiltonian is
\bea
H(t)&=&\int d^dx\;[{\8{\it \Pi}}^\nu(t,x)\pa_{x^0}\8 \CA_\nu(t,x)-
\D(x^0)\CL(t,x)],
\label{Ham}
\eea
where ${\8{\it \Pi}}^\nu$ is a momentum for $\8 \CA_\nu$ and 
\bea
\CL(t,x)&=&-\frac14~\8 \CF_{\mu\nu}(t,x)\8 \CF^{\mu\nu}(t,x),
\\
\8 \CF_{\mu\nu}(t,x)&=&\pa_\mu \8 \CA_\nu(t,x)-
\pa_\nu \8 \CA_\mu(t,x)-i
[\8 \CA_\mu(t,x),\8 \CA_\nu(t,x)].
\eea
Note that using \bref{Qdot}, now the
{\it star} product is defined with respect to $x^\mu=(\s,\bx)$ 
instead of $x^\mu=(t,\bx)$ in \bref{Moyal}. Thus it 
contains spatial derivatives of infinite order but no 
{time} derivative. The same applies for the Hamiltonian, it contains 
no derivative with respect to $t$, and so it is a good phase-space quantity,
a function of the canonical pairs 
$(\8 \CA_\mu(t,x),\8 {\it\Pi}^\mu(t,x))$ with Poisson bracket
\bea
\{\8 \CA_\mu(t,x),\8 {\it\Pi}^\nu(t,x')\}&=&
{\D_\mu}^\nu~ \D^{(d)}(x-{x}'). 
\eea
The momentum constraint  \bref{vp} is 
\bea
\vp^\nu(t,x)&=&\8 {\it\Pi}^\nu(t,x)+\int dy~\chi(x^0,-y^0)\;
\8 \CF^{\mu\nu}(t,y)\;\8 \CD^y_\mu\D(x-y)
\nn\\
 &=&
\8 {\it\Pi}^\nu(t,x)+\D(x^0)\8 \CF^{0\nu}(t,x)-\frac{i}{2}
\left(\ep(x^0)[\8 \CF^{\mu\nu},\8 \CA_\mu]
-[\ep(x^0)\8 \CF^{\mu\nu},\8 \CA_\mu]\right)\approx 0.
\nn\\
\label{vpu1}
\eea
while the constraint \bref{EOM}, obtained from the consistency of
the above one, turns out to be 
\bea
\7\vp^\nu(t,x)&=&\8 \CD_\mu \8 \CF^{\mu\nu}(t,x)\approx 0.
\label{EOMu1}
\eea
Note that these constraints are reducible since $\8\CD_\mu\7\vp^\mu\equiv 0$.
They reproduce the EL equation of motion \bref{ELeq} 
using the Hamilton equation \bref{Qdot},
\bea
\pa_{t}{\8\CA}_\mu(t,x)&=&\{\8 \CA_\mu(t,x), H(t)\}~=~
\pa_{x^0}\8 \CA_\mu(t,x)
\eea
and the identification \bref{qdef}, $\8 \CA_\mu(t,x^\nu)=\8 
A_\mu(t+x^0,\bx)$.
Since the \lag of \bref{Maxwell} has translational invariance,
the Hamiltonian \bref{Ham}, as well as the 
constraints \bref{vpu1} and \bref{EOMu1}, are conserved.

To compute the generator of the {$U(1)$} \NC transformation,
we apply \bref{G} to our case
\bea
G[\8\Lam]&=&\int dx[\;\8 {\it\Pi}^\mu \D \8 \CA_\mu-\D(x^0)\;\cf^0\;],
\label{defG}
\eea
where the last term must be evaluated from surface term appearing
in the variation of the Lagrangian  
\bea
&&\int dx[-\D(x^0)\;\cf^0\;]=
\int dx[\frac{\ep(x^0)}{2}\;\pa_{\mu}\cf^\mu\;]=
\int dx[\frac{\ep(x^0)}{2}\;\D \CL\;].
\label{defFF}
\eea
Using \bref{Dlag} the $U(1)$ generator becomes
\bea
G[\8\Lam]&=&\int dx\left[\8 {\it\Pi}^\mu \8 \CD_\mu\8\Lam~+~
\frac{i}{4}{\ep(x^0)} \8 \CF_{\mu\nu}~[\8 \CF^{\mu\nu},\8\Lam]\right],
\label{defG2}
\eea
where, as discussed in \bref{Lamdd}, $\8\Lam(t,x^\mu)$ must be an arbitrary 
function satisfying
\bea
\dot{\8\Lam}(t,x^\mu)&=&\pa_{x^0}\8\Lam(t,x^\mu)
\label{Lamdd1}
\eea
The generator can be expressed as a linear combination of the constraints, 
\bea
G[\8\Lam]&=&\int dx~\8\Lam\left[-(\8 \CD_\mu\vp^\mu)-\D(x^0)\7\vp^0+
\frac{i}{2}\left(\ep(x^0)[\7\vp^\nu,\8 \CA_\nu]-[\ep(x^0)\7\vp^\nu,
\8 \CA_\nu]
\right)\right].
\nn\\
\label{GLam}
\eea
The fact that the generator \bref{GLam} is a sum of constraints
shows explicitly
the conservation of the generator on the constraint surface.
It also means the $U(1)$ invariance of the Hamiltonian 
on the constraint surface.
Furthermore $G[\8\Lam]$ is conserved, without using constraints, 
for $\8\Lam(t,x)$
satisfying \bref{Lamdd1},
\bea
\frac{d}{dt}G[\8\Lam]&=&\{G[\8\Lam],H\}~+~\frac{\pa}{\pa t}G[\8\Lam]~=~0
\label{dtGLam}
\eea
in agreement with \bref{dtG}.

\vskip 4mm

Finally, the Hamiltonian turns out to be
\bea
H&=&G[\8\CA_0]\;+\;\int dx\;\vp^i\;\8\CF_{0i}\;+\;E_L,
\eea
where the first term is the $U(1)$ generator \bref{GLam}
replacing the parameter $\8\Lam$ by $\8\CA_0$. The last term $E_L$ is 
the only relevant one on the constraint surface, and it is
\bea
E_L&=&
\int dx\; \D(x^0)\;\{\frac12 {\8\CF_{0i}}^2+\frac14 {\8\CF_{ij}}^2 \}
\nn\\
&& + 
\frac{i}{2}\int d x\;\8\CA_0\left( 
\frac12 [\8\CF^{ij}, \ep(x^0)\8\CF_{ij}]-[ \8\CF^{0i},\ep(x^0)
\8\CF_{0i}]\right)
\nn\\&&+
\frac{i}{2}\int d x\;\8\CA_j\left( 
[\8\CF_{0i}, \ep(x^0)\8\CF^{ij}]-[\ep(x^0) \8\CF_{0i},\8\CF^{ij}]\right).
\eea
This expression is useful, for example, 
to evaluate the energy of classical configurations of the theory.
The two terms in the first line have the same form as the "energy" of
the commutative $U(1)$ theory. The last two lines are non-local contributions.
However they vanish in two cases,  (1) in $\T^{0i}=0$ (magnetic) 
background
and 
(2) for $t$ independent solutions of $\CA_\mu$. 

 \vskip 6mm

\section{Seiberg-Witten map, gauge generators and Hamiltonians}
\indent

Seiberg and Witten \cite{sw} have introduced a map between the gauge
potential $A_\mu$ in a $U(1)$ {commutative} and $\8 A_\mu$ in an $U(1)$ 
{\NC} theories. Here we show that the Seiberg-Witten (SW) map for the 
space-time  $U(1)$ \NC theories can be viewed as a {\it canonical transformation}
in the \ham formalism in $\dpo$ dimensions. This makes
explicit the physical equivalence of both theories.
By finding the corresponding
generating functional, we are able to map quantities between theories.
In particular, we show how the gauge generator and the \ham obtained
in the previous section for the $NC$ case are mapped to those of the
commutative theory.

\subsection{The $d$ formalism}

We recall that the SW map from the $U(1)$ commutative
connection $A_\mu$ to the $U(1)$ \NC one $\8A_\mu$ looks like
\bea
\8 A_\mu=A_\mu+\frac12\T^{\r\s} A_\s(2\pa_\r A_\mu-\pa_\mu A_\r)
+...\,.
\label{Ahat}
\eea
In the following discussions we keep terms only up to the first order in 
$\T$ and higher power terms of $\T$,  indicated by $...$, are omitted.

Under a commutative $U(1)$ transformation of $\D A_\mu=\pa_\mu\lam$, the
mapped $\8 A_\mu$ transforms as
\bea
\D\8 A_\mu&=&
\pa_\mu\{\lam
+\frac12\T^{\r\s} A_\s\pa_\r\lam
\}+\T^{\r\s} \pa_\s\lam\pa_\r A_\mu~
=~\8 D_\mu\8\lam.
\label{DAhat}
\eea
Note that although the field $\8A_\mu$ defined above transforms as  $U(1)$ 
\NC gauge potentials the gauge transformation parameter 
function $\8\lam$ is now gauge field dependent
\bea
\8\lam(\lam,A)=\lam+\frac12\T^{\r\s} A_\s\pa_\r\lam.
\label{lamhat}
\eea
The field strength $\8 F_{\mu\nu}$ defined as in \bref{defF} is,
in terms of the commutative fields $A_\mu$ and 
$F_{\mu\nu}\equiv\pa_\mu A_\nu-\pa_\nu A_\mu$, as
\bea
\8 F_{\mu\nu}=F_{\mu\nu}+\T^{\r\s} F_{\r\mu}F_{\s\nu}-
\T^{\r\s}A_{\r}\pa_\s F_{\mu\nu}
\eea
and transforms under  $\D A_\mu=\pa_\mu\lam$ covariantly as
\bea
\D \8 F_{\mu\nu}=-\T^{\r\s} \pa_{\r}\lam\pa_\s F_{\mu\nu}
=-i[ F_{\mu\nu},\lam]=-i[\8 F_{\mu\nu},\8\lam].
\eea
\medskip 

\subsection{The $d+1$ formalism}

In the  $\dpo$ dimensional \ham formalism 
 we can regard the mapping \bref{Ahat} as a canonical 
transformation. 
Denoting the $\dpo$ dimensional potentials  
$\8\CA_\mu(t,x)$ and $\CA_\mu(t,x)$ corresponding to $d$ dimensional
ones $\8A_\mu(t,\bx)$ and $A_\mu(t,\bx)$ respectively
\footnote{Remember, hats for fields in the non-commutative theory,
and calligraphic letters for fields in the $d+1$ formalism},
the generating function turns out to be 
\bea  
W(\CA,\8{\it\Pi})=\int d^dx\;\8{\it\Pi}^\mu\left(
\CA_\mu+\frac12\T^{\r\s} \CA_\s(2\pa_\r \CA_\mu-\pa_\mu \CA_\r)
\right)\;+\;W^0(\CA),
\label{GF}
\eea
where $W^0(\CA)$ is any function of $\CA_\mu$ of order $\T$.
It generates the transformation of $\CA_\mu$ as in \bref{Ahat} 
\bea  
\8\CA_\mu&=&
\CA_\mu+\frac12\T^{\r\s} \CA_\s(2\pa_\r \CA_\mu-\pa_\mu \CA_\r)
\label{hAhat}
\eea
and determines the relation between ${\it\Pi}^\mu$ and $\8{\it\Pi}^\mu$,
conjugate momenta of $\CA_\mu$ and $\8 \CA_\mu$ respectively, to be
\bea
{\it\Pi}^\mu=\8{\it\Pi}^\mu+\frac12\8{\it\Pi}^\s\T^{\r\mu} 
(2\pa_\r \CA_\s-\pa_\s \CA_\r)-\pa_\r(\T^{\r\s} \CA_\s\8{\it\Pi}^\mu)
+\frac12 \pa_\r (\8{\it\Pi}^\r\T^{\mu\s}\CA_\s)+
\frac{\D W^0(\CA)}{\D \CA_\mu}.
\nn\\
\eea
It can be inverted, to first order in $\T$, as
\bea
\8{\it\Pi}^\mu={\it\Pi}^\mu+\T^{\mu\r}{\it\Pi}^\s \CF_{\r\s}+
{\it\Pi}^\mu\frac12\T^{\r\s}\CF_{\r\s}+\T^{\r\s} \CA_\s\pa_\r {\it\Pi}^\mu
-\frac12 (\pa_\r {\it\Pi}^\r)\T^{\mu\s}\CA_\s-
\frac{\D W^0(\CA)}{\D \CA_\mu}.
\label{Pihat}
\nn\\
\eea
Note that the canonical transformation, \bref{hAhat} and \bref{Pihat}, 
is independent of the concrete theories we are considering.
\medskip

In the last section the generator of $U(1)$ \NC theory
was obtained in \bref{defG2} as
\bea
G[\8\Lam]&=&\int dx\left[\8{\it\Pi}^\mu \8\CD_\mu\8\Lam~+~
\frac{i}{4}{\ep(x^0)} \8\CF_{\mu\nu}~[\8\CF^{\mu\nu},\8\Lam]\right].
\label{defG22}
\eea
The last term appeared since the original Lagrangian $L^{non}$ 
changes to a surface term as in \bref{Dlag}
under the gauge transformation.
Now we want to see how this generator transforms under the SW map.
It is straightforward to show that, for $W^0(\CA)=0$, 
\bea
\int dx[~\8 {\it\Pi}^\mu \8 \CD_\mu\8\Lam(\Lam,\CA)~]
&=& 
\int dx~[\; {\it\Pi}^\mu \pa_\mu\Lam\;],
\label{defGc}
\eea
where 
\bea
\8\Lam(\Lam,\CA)=\Lam+\frac12\T^{\r\s} \CA_\s\pa_\r\Lam,~~~~~~~~~~~
\dot \Lam=\pa_{x^0}\Lam.
\label{lamhat2}
\eea
These results are
independent of the specific form of Lagrangian for $U(1)$
\NC and commutative gauge theories.
On the other hand the term $\delta(\sigma) \cf(t,\s)$ appearing in \bref{G}
does depend on the specific theory we are considering.
For the $U(1)$ \NC  theory, \bref{Maxwell},
it is  nothing but the \lag 
dependent term in \bref{defG22}, which expanding to first order in $\T$  
\bea
\frac{i}{4}\int dx\;{\ep(x^0)} \8 \CF_{\mu\nu}[\8 \CF^{\mu\nu},\8\Lam]\;
=~
\frac{1}{4}\int dx\;{\D(x^0)}\T^{0i}
  \CF_{\mu\nu} \CF^{\mu\nu}\pa_i\Lam.
\eea
In this case the generator of $U(1)$ \NC transformations
can be mapped to that of commutative one
\bea
G[\8\Lam(\Lam,\CA)]&=&
\int dx\; \{{\it\Pi}^0 \pa_0\Lam\;+({\it\Pi}^i+
\frac{1}{4}{\D(x^0)}\T^{0i}
  \CF_{\mu\nu} \CF^{\mu\nu})\pa_i
\Lam\}-\int dx \frac{\D W^0(\CA)}{\D \CA_\mu}\pa_\mu\Lam
\nn\\&=&
\int dx\; [~{\it\Pi}^\mu \pa_\mu\Lam~]
\label{Gcom}
\eea
if we choose the canonical transformation with
\bea
W^0(\CA)&=&\frac14\int d x~\D(x^0)~\T^{0\mu}\CA_\mu \CF_{\r\s}\CF^{\r\s}.
\label{W00}
\eea
The right hand side of \bref{Gcom} is the well-known generator of the 
$U(1)$ commutative theory (see the appendix).

Now we would like to see what is the form of the $U(1)$ Hamiltonian
obtained from \bref{Ham} under the SW map, \bref{hAhat} and \bref{Pihat}. 
The $U(1)$ commutative Hamiltonian results to be 
\be
H^{(c)}
=\int dx\;[{{\it \Pi}}^\nu(t,x) \CA'_\nu(t,x)~-~
\D(x^0) \CL^{(c)}(t,x)]\ee
\label{abelianham}
where
\be
\CL^{(c)}(t,x)= -\frac14 \CF^{\nu\mu}\CF_{\nu\mu}-\frac12\CF^{\mu\nu}
\T^{\r\s} \CF_{\r\mu}\CF_{\s\nu}+
\frac18 \T^{\nu\mu}\CF_{\nu\mu}\CF_{\r\s}\CF^{\r\s}.
\label{U1lag}
\ee
But this is nothing but the $ \dpo$ dimensional Hamiltonian 
that we would have obtained from an abelian $U(1)$ gauge theory with \lag
\bea
L^{(c)}(t,\bx)&=&
-\frac14 F^{\nu\mu}F_{\nu\mu}-
\frac12 F^{\mu\nu}\T^{\r\s} F_{\r\mu}F_{\s\nu}+
\frac18 \T^{\nu\mu}F_{\nu\mu}F_{\r\s}F^{\r\s}
\label{U1BIa}
\eea
in $d$ dimensions. 
One can check that this Lagrangian is, up to a total derivative, the
expansion of the Born-Infield action up to order $F^3$,
when written in terms of the open string parameters \cite{sw}
\footnote{We acknowledge discussions with Joan Sim\'on on this
point.}.
\bea
L^{(c)}&\sim&1-\sqrt{-\det(\h_{\mu\nu}-\T_{\mu\nu}+F_{\mu\nu})}~\sim~
1-\sqrt{-\det(\h_{\mu\nu}+\8F_{\mu\nu})}.
\label{relBI}
\eea

\vskip 6mm

\section{BRST symmetry}
\indent

In this section we will conclude our work with the $U(1)$ $NC$
gauge theory by studying its BRST and field-antifield properties.
First of all, we will study the BRST symmetry \cite{brs}\cite{t} at
classical and quantum levels. We will construct 
the BRST charge and the BRST invariant Hamiltonian 
working with the $\dpo$ dimensional formulation, and we will
check the nilpotency of the BRST generator.
Then, in order to map the BRST charges and Hamiltonians 
of the $U(1)$ \NC  and commutative
$U(1)$ gauge theories, we will generalize the
SW map to the superphase space.

Finally, in the last subsection,
we will also study the BRST symmetry at \lag level using the
field-antifield formalism \cite{z}\cite{bv}, 
for a review see \cite{ht}\cite{gps}\cite{w}.
We will construct the solution of the
classical master equation in the classical and gauge fixed basis. 
We will also realize the SW map as an antibracket canonical
transformation.

\subsection{Hamiltonian BRST charge}
\indent

The BRST symmetry at classical level encodes the classical
gauge structure through the nilpotency of the BRST transformations of
the classical fields and ghosts \cite{ku}\cite{bv85}\cite{fh90}.
The BRST symmetry of the classical fields is constructed from the
gauge
transformation by changing the gauge parameters by ghost fields. 

Let us consider again the $U(1)$ $NC$ theory still in $d$ dimensions.
Its BRST transformations are
\bea
\D_B\8  A_\mu&=&\8 D_\mu \8 C,~~~~~\D_B \8 C~=~-i \8 C*\8 C,
\\
\D_B \8 {\ba C}&=&\8 B,~~~~~~~~~\D_B \8 B~=~0,
\eea
where $\8 C, \8{\ba C}, \8B $ are the ghost, antighost and auxiliary field 
respectively.

These are again a symmetry of the \lag associated 
with \bref{Maxwell}, since its change under the BRST 
transformations is
\be
\D_B L~=~ 
\frac{i}{2}~[\8 F_{\mu\nu},\8 C]~\8 F^{\mu\nu}.
\label{Dlagc}
\ee
which, as in \bref{Dlag}, can be shown a total divergence.
We can construct the gauge fixing \lag
$\8L_{gf+FP}$ by adding the proper term of the form $\D_B\8\Psi$.
In this case, the gauge fixing fermion is 
\be
\8 \Psi= \8{\ba C}~(\partial^\mu \8 A_\mu +\alpha \8 B)
\label{GFF}
\ee
Then the $\8 L_{gf+FP}$ is, up to a total derivative,
\bea
\8 L_{gf+FP}&=&-~
\partial^\mu\8{\ba C}~ \8 D_\mu \8 C~+~
\8 B~(\partial^\mu \8 A_\mu +\alpha \8 B).
\eea
By construction, this term does not spoil the symmetry. Indeed 
\bea
\D_B \8 L_{gf+FP}&=& 
\pa^\mu(\8 B \8 D_{\mu}\8 C).
\label{Dlaggfc}
\eea 

\vskip 6mm

In order to construct the generator of the BRST transformations and the
BRST invariant Hamiltonian we should use the
$\dpo$ dimensional formulation.
We denote the $\dpo$ dimensional fields 
corresponding to the  $d$ dimensional ones $\8 C, \8{\ba C}, \8B $,
using with the calligraphic letters, as $\8\CC, \8{\ba \CC}, \8\CB $
respectively. The results are that the BRST invariant Hamiltonian is given by
\bea
H(t)&=&H^{(0)}~+~H^{(1)}
\label{Hamc01}\\
H^{(0)}&=&\int dx\;[{\8{\it \Pi}}^\nu(t,x)\8 \CA'_\nu(t,x)~+~
{\8{ \CP_c}}(t,x)\8 \CC'(t,x)~-~\D(x^0)\8 \CL^0(t,x)],
\\
H^{(1)}&=&\int dx\;[{\8{\CP}}_\CB\8 \CB'(t,x)~+~
{\8{ \CP_{\ba \CC}}}(t,x)\8{\ba \CC}'(t,x)~-~
\D(x^0)\8 \CL_{gf+FP}(t,x)].
\label{Hamc2}
\eea
while the  BRST charge is
\bea
Q_B&=&Q_B^{(0)}~+~Q_B^{(1)}
\label{BRSTQ}\\
Q_B^{(0)}&=&\int dx\left[\8 {\it\Pi}^\mu \8 \CD_\mu\8\CC~-i~\8{ \CP_\CC}*
\8\CC*\8\CC~+~\frac{1}{2}{\ep(x^0)} \D_B \8 \CL^0(t,x)\right].
\\
Q_B^{(1)}&=&\int dx\left[~\8{ \CP_{\ba \CC}} ~\8 \CB~+~
\frac{1}{2}{\ep(x^0)} \D_B \8 \CL_{gf+FP}(t,x)\right],
\label{defG2c}
\eea
It is  an analogue of the BFV charge \cite{fv75}\cite{bv77}
 for $U(1)$  \NC theory.
$H^{(0)}$ , $Q_B^{(0)}$ are the "gauge unfixed" and 
 the $H$, $Q_B$ are "gauge fixed" Hamiltonians and BRST charges.

Using the graded symplectic structure of the superphase space \cite{roberto}
\bea
\{\8\CA_\mu(t,x),\8{\it\Pi}^\nu(t,x')\}&=&{\D_\mu}^\nu~ \D^{(d)}(x-{x}')
,~~~
\{\8{\cal {C}}(t,x),\8{\cal P}_{\8{\cal {C}}}(t,x')\}~=~
\D^{(d)}(x-{x}'),
\nn\\
\{\8{{\ba \CC}}(t,x),\8{\cal P}_{\8{\ba{\CC}}}(t,x')\}&=&
\D^{(d)}(x-{x}'),~~~~~~~~
\{\8{\CB}(t,x),\8{\cal P}_{\8{\CB}}(t,x')\}~=~
\D^{(d)}(x-{x}')
\nn\\ \eea
we have
\bea
\{H^{(0)},Q_B^{(0)}\}~=~\{Q_B^{(0)},Q_B^{(0)}\}~=~0,
\eea
and
\bea
\{H,Q_B\}~=~\{Q_B,Q_B\}~=~0.
\eea
Thus the BRST charges are nilpotent and the Hamiltonians are 
BRST invariant both in the gauge unfixed and the gauge fixed 
levels.


\subsection{Seiberg-Witten map in superphase space}
\indent

Now we would like to see how the BRST charges and the 
BRST invariant Hamiltonians
of the \NC and commutative gauge theories are related. 
In order to do that we will extend the SW map to a 
canonical transformation in the superphase space
$(\CA,\CC,{\ba{\cal {C}}},\CB,\iPi,{\cal P}_{\cal {C}},
\CP_{\ba{\cal {C}}},\CP_\CB)$.  We introduce the generating 
function 
\bea  
W(\CA,\CC,{\ba{\cal {C}}},\CB,\8\iPi,\8{\cal P}_{\cal {C}},
\8\CP_{\ba{\cal {C}}},\8\CP_\CB)
&=&\int dx\;\left[\8\iPi^\mu\left(
\CA_\mu+\frac12\T^{\r\s} \CA_\s(2\pa_\r \CA_\mu-\pa_\mu \CA_\r)
\right)\;\right.
\nn\\&&\left.+\;\8{\cal P}_{\cal {C}}
\left(\CC+\frac12\T^{\r\s} \CA_\s\pa_\r \CC \right)~+~
\8{\cal P}_{\ba{\cal {C}}}{\ba{\cal {C}}}~+~
\8{\cal P}_{\CB}{\CB}\right]
\nn\\&&~+~
W^0(\CA,\CC)~+~W^1(\CA,\CC,{\ba{\cal {C}}},\CB),
\nn\\
\label{GF3}
\eea
where $W^0(\CA,\CC)$ depends on the 
specific form of the $U(1)$ \NC Lagrangian and  

\noindent
$W^1(\CA,\CC,{\ba{\cal {C}}},B)$
also on the form of the gauge fixing. For the $U(1)$ 
\NC theory and for the
gauge fixing \bref{GFF}, we have 
\bea
W^0(\CA,\CC)&=&\frac14\int d x~\D(x^0)~\T^{0\mu}\CA_\mu \CF_{\r\s}
\CF^{\r\s}
\label{W0}
\eea
as in \bref{W00} and
\bea
W^1&=&\int dx ~\frac{1}{2}{\ep(x^0)}
\left[\pa^\mu \{\frac12\T^{\r\s}\CA_\s(2\pa_\r\CA_\mu-\pa_\mu\CA_\r)\}
\CB\right.
\nn\\&&~+~\left.
\{\frac12\T^{\r\s}\CA_\s(2\pa_\r\CA_\mu-\pa_\mu\CA_\r)\pa_\s\CC+
\frac12\T^{\r\s}\CA_\s\pa_\mu\pa_\r{\cal {C}}~\}~\pa^\mu\ba\CC~\right].
\label{solw2}
\eea
The transformations are obtained from the generating function by 
\bea
\8\Phi^A=\frac{\partial_\l W}{\partial \8 P_A},~~~~~\quad
P_A=\frac{\partial_r W}{\partial \Phi^A},
\eea
where $\Phi^A$ represent any fields, $P_A$ their conjugate momenta,
and $\pa_r$ and $\pa_\l$ are right and left derivatives respectively.

Explicitly we have
\bea
\8 \CA_\mu&=&\CA_\mu+\frac12\T^{\r\s} \CA_\s(2\pa_\r \CA_\mu-
\pa_\mu \CA_\r),
\\
\8{\cal {C}}&=&{\cal {C}}+\frac12\T^{\r\s}\CA_\s\pa_\r{\cal {C}},
\label{chata}
\\
\8{\ba\CC}&=&{\ba\CC},
\\
\8\CB&=&\CB,~~~~~~~~~~~~~
\eea
and
\bea
\8\iPi^\mu&=&\iPi^\mu+\T^{\mu\r}\iPi^\s \CF_{\r\s}+
\iPi^\mu\frac12\T^{\r\s}\CF_{\r\s}
+\T^{\r\s} \CA_\s\pa_\r \iPi^\mu
-\frac12 (\pa_\r \iPi^\r)\T^{\mu\s}\CA_\s
\nn\\
&&~~+\frac12{\cal P}_{\CC}\T^{\mu\s}\pa_{\s}\CC~-~
\frac{\D (W^0+W^1)}{\D \CA_\mu},
\\
\8{\cal P}_{\CC}&=&{\cal P}_{\CC}+\frac12\T^{\r\s} \partial_\rho 
({\cal P}_{\CC}\CA_\s)~-~
\frac{\D_r (W^0+W^1)}{\D \CC},
\\
\8{\cal P}_{\ba\CC}&=&{\cal P}_{\ba\CC}~-~
\frac{\D_r W^1}{\D {\ba\CC}},
\\
\8{\cal P}_{\CB}&=&{\cal P}_{\CB}~-~
\frac{\D_r W^1}{\D \CB}.
\eea
Using this transformation we can rewrite the BRST charge \bref{BRSTQ}
as 
\bea
Q_B&=&Q_B^{(0)}~+~Q_B^{(1)}~=~
\int dx [\iPi^\mu  \pa_\mu\CC~+~{ \CP_{\ba \CC}}  \CB~-~
\D(x^0) \CB\pa^0 \CC~]
\nn\\&=&
\int dx [\iPi^\mu  \pa_\mu\CC~+~{ \CP_{\ba \CC}}  \CB~+~
\frac{1}{2}{\ep(x^0)} \D_B \CL_{gf+FP}(t,x)~],
\label{totQBgf}
\eea
 where $\CL_{gf+FP}(t,x)$ is the abelian gauge fixing
Lagrangian and is given by
\bea
\CL_{gf+FP}&=&-\partial^\mu\overline {\CC}~ \pa_\mu  \CC~+~
\CB~(\partial^\mu  \CA_\mu +\alpha  \CB).
\label{laggf0}
\eea 
 The total $U(1)$  \ham \bref{Hamc01} becomes 
\bea
H&=&\int dx\;[{{\it \Pi}}^\nu  \CA'_\nu +{{ \CP_\CC}} \CC'  
+{{ \CP_{\ba\CC}}} {\ba\CC}'+{{\it P}}_\CB \CB'
-\D(x^0) (\CL^{(c)}+\CL_{gf+FP}) ].
\label{totHam}
\eea
Remember $\CL^{(c)}$ is the $U(1)$ commutative \lag given in \bref{U1lag}.
Summarizing, we have been successful in mapping the $NC$ and commutative
charges in the $d+1$ formalism by generalizing the SW map to a
canonical transformation in the superphase space.

\vskip 6mm
\subsection{ Field-antifield formalism for $U(1)$  non-commutative theory}
\indent

The field-antifield formalism allows us to study the BRST symmetry 
of a general gauge theory by introducing a canonical
structure at a Lagrangian level \cite{z}\cite{bv}\cite{ht}\cite{gps}.
The classical master equation in the classical basis encodes the 
gauge structure of the generic gauge theory \cite{bv85}\cite{fh90}.
The solution of the classical master equation in the gauge fixed basis
gives the ``quantum action'' to be used in the path integral quantization. 
Any two solutions of the classical master equations
are related by a canonical transformation in the antibracket sense \cite{vt}.

Here we will apply these ideas to the $U(1)$  \NC theory.
Since we work at a \lag level we will work in $d$ dimensions. 
In the classical basis the set of fields and antifields are
\be
\Phi^A=\{ \8 A_\mu, \8 C\},~~~~~~~ \Phi^*_A=\{ \8 A^*_\mu,\8 C^* \}.
\ee
The solution of the classical master equation 
\be
(S,S)=0,
\ee
is given by{\footnote{
As in usual convention in the antifield formalism, $d$ dimensional
integration is understood in summations.}}
\be
S[\Phi,\Phi^*]=I [\8 A]+\8 A^*_\mu \8 D^\mu \8 C-i\8 C^* (\8 C * \8 C),
\ee
where $I [\8 A]$ is the classical action and the antibracket 
$ (~~,~~)$ is  defined by 
\be
(X,Y)=\frac {\partial_r X}{\partial \Phi^A} 
\frac {\partial_l Y}{\partial \Phi^*_A}-
\frac {\partial_r X}{\partial \Phi^*_A} 
\frac {\partial_l Y}{\partial \Phi^A}.
\ee

The gauge fixed basis can be analyzed by introducing  the antighost
and auxiliary fields and the corresponding antifields. It can be
obtained from the classical basis by considering a canonical transformation, in
the antibracket sense,
\bea
\Phi^A&\longrightarrow& \Phi^A
\nn\\
\Phi^*_A&\longrightarrow& \Phi^*_A +\frac {\partial_r \Psi}
{\partial \Phi^A}
\eea
generated by
\be
\8\Psi=  \8{\ba C}~(\partial^\mu \8 A_\mu +\alpha \8 B),
\ee
where $\8{\ba C}$ is the antighost and $\8 B$ is the auxiliary field.
We have 
\bea 
S[\Phi,\Phi^*]=\8I_\Psi +\8 A^{*\mu} \8 D_\mu \8 C- i\8 C^* (\8 C * \8 C)
+{\8 {\ba C}}^* \8 B,
\label{SPP}
\eea
where $\8I_\Psi$ is the ``quantum action'' and is given by
\bea
\8I_\Psi= I[\8 A]+(-\partial_\mu \8{\ba C}~\8D^\mu\8 C+
\8 B~\partial_\mu\8 A^\mu+\alpha\8 B^2).
\label{IPSI}\eea
The action $\8I_\Psi$ has well defined propagators and is the starting
point of the Feynman perturbative calculations.
\vspace{3mm}

Now we would like to study what is the SW map in the space of
fields and antifields. We first consider it in the classical basis.
 In order to do that we construct a canonical
transformation in the antibracket sense
\bea
\8\Phi^A=\frac{\partial_l F_{cl}
[\Phi,\8 \Phi^*]}{\partial \8 \Phi^*_A},\quad
\Phi^*_A=\frac{\partial_r F_{cl}[\Phi,\8 \Phi^*]}{\partial \Phi^A},
\eea
where 
\bea
F_{cl}
=\8 A^{*\mu} \left(A_\mu+\frac12\T^{\r\s} A_\s(2\pa_\r A_\mu-\pa_\mu A_\r)
\right)+\8 C^*(C+\frac12\T^{\r\s} A_\s\pa_\r C).
\label{FCL}\eea
The gauge structures of  \NC and commutative 
are mapped to each other
\be
 \8 A^*_\mu \8 D^\mu \8 C- i\8 C^* (\8 C * \8 C)=A^*_\mu\partial^\mu C.
\ee

We can generalize the previous results to the gauge fixed
basis. In this case the transformations of the antighost 
and the auxiliary field sectors should be taken into account. 
The generator of the 
canonical transformation is modified from \bref{FCL} to
\bea
F_{gf}
&=&F_{cl}~+~
\left(\8{\ba C}^*+\frac12\T^{\r\s}\pa^\mu \left(
A_\s(2\pa_\r A_\mu-\pa_\mu A_\r)\right)
\right)\overline {C}+\8B^* B.
\label{SWgf}
\eea
Note that the additional term gives rise to new terms in $A^{*\mu}$ and 
$\overline {C}^*$ 
while the others remain the same as in the classical basis.
In particular 
\bea
\8{\overline {C}}&=&\overline {C},~~~~~~\8B~=~B.
\eea
Using the transformation we can express  \bref{SPP} and \bref{IPSI} as
\bea 
S[\Phi,\Phi^*]=I_\Psi + A^{*\mu} \pa_\mu C+{\overline{C}}^*  B
\label{SPPgf}
\eea
where 
\bea
I_\Psi= I[\8 A(A)]+ (-\partial_\mu \overline{ C}~\pa^\mu C+
B~\partial_\mu A^\mu+\alpha B^2)
\label{IPSIgf}
\eea
and $I[\8 A(A)]$ is the classical action in terms of $A_\mu$.
This is indeed a quantum action for the commutative $U(1)$ BRST
invariant action in the gauge fixed basis.
In this way the canonical transformation \bref{SWgf} 
maps the  $U(1)$ \NC structure of the $S[\Phi,\Phi^*]$ into the 
commutative one in the gauge fixed basis.

\vskip 6mm

\section{Discussions}
\indent

 In this paper the \ham formalism of the non-local theories is 
discussed by using $\dpo$ dimensional formulation \cite{lv}\cite{gkl}. 
For a given non-local \lag in $d$ dimensions the Hamiltonian is 
introduced by \bref{h} on the phase space of the $\dpo$ dimensional fields. 
The equivalence with the original non-local theory is 
assured by imposing two constraints \bref{vp} and \bref{EOMf} consistent 
with the time evolution. 
The degrees of freedom of the extra dimension (denoted by coordinate $\s$) 
has its origin in the infinite degrees of freedom associated with the 
non-locality.
The fact that we have been led to a theory with ``two times''
should be intimately related to their acausality \cite{sst}\cite{ab} and
non-unitarity \cite{gm}\cite{agm}. 

The $d+1$ formalism is also applicable to {\it local} and 
higher derivative theories. 
In these cases the set of
constraints are used to reduce the redundant degrees of freedom
of the infinite dimensional phase space,
reproducing the standard $d$ dimensional formulations \cite{gkll2}.    

We have analyzed the symmetry generators of non-local 
theories in the \ham formalism. 
As an example we have considered the space-time $U(1)$ \NC 
gauge theory. 
The gauge transformations in $d$ dimensions are
described as a rigid symmetry in $\dpo$ dimensions. 
The generators of 
{\it rigid} transformations in $\dpo$ dimensions turn out to be the 
generators of {\it gauge}
transformations when the reduction to $d$ dimensions can be performed
as is shown for the $U(1)$ commutative gauge theory in the appendix.  

We have extended the Seiberg-Witten map  to a canonical transformation.
This allows us to map the Hamiltonians and the 
gauge generators of non-commutative and 
commutative theories. We have also seen explicitly the map of the 
$U(1)$ \NC and the BI actions up to $F^3$.  
The reason why we were able to discuss the SW map as a canonical 
transformation is that
we have considered the phase space of the commutative theory also
in the $\dpo$ dimensions.

The BRST symmetry has been analyzed at Hamiltonian and Lagrangian levels.
The relation between the $U(1)$ commutative and  \NC parameter functions 
is understood as a canonical transformation of the ghosts 
in the super phase space of the SW map. 
Using the field-antifield formalism we have seen how the solution of the
classical master equation for non-commutative and 
commutative theories are related by a canonical transformation in the 
antibracket sense. This results shows that the antibracket cohomology
classes of both theories coincide in the space of non-local functionals.
The explicit forms of the antibracket canonical transformations could be
useful to study the observables, anomalies, etc. in the $U(1)$ \NC theory.

\medskip

\vskip 6mm

{\it{\bf Acknowledgments}}

We acknowledge discussions with Luis Alvarez-Gaum\'e, Josep Llosa, 
David Mateos,  Josep Mar\'{\i}a Pons, Joan Sim\'on and Fredy Zamora.
K.K. would like to thank CERN for their hospitality during his stay. 
The work of J.G is partially supported by AEN98-0431, GC 1998SGR (CIRIT) 
and K.K. by the Grant-in-Aid for Scientific Research, 
No.12640258 (Ministry of Education Japan).
T.M. is supported by a fellowship from the Commissionat per 
a la Recerca de la Generalitat de Catalunya.

\vskip 6mm

\appendix
\section{$U(1)$ commutative Maxwell theory in $\dpo$ dimensions}
\indent

Our $d+1$ formalism can also be used for describing ordinary 
local theories. As an example of this, we will show
how the $U(1)$ {commutative} 
Maxwell theory is formulated using
the $\dpo$ dimensional canonical formalism developed for non-local 
theories in section 2 and see
how it is reduced to the standard canonical formalism 
in $d$ dimensions. 

The canonical $d+1$ system 
is defined by the Hamiltonian \bref{h} and two constraints,
 \bref{vp} and \bref{EOM}.
The Hamiltonian is
\bea
H&=&\int d^dx\;[{\it\Pi}^\nu(t,x)\pa_{x^0}\CA_\nu(t,x)-\D(x^0)\CL(t,x)~],
\label{Hama}
\eea
where 
\bea
\CL(t,x)&=&-\frac14~\CF_{\mu\nu}(t,x)\CF^{\mu\nu}(t,x),
\\
\CF_{\mu\nu}(t,x)&=&\pa_\mu \CA_\nu(t,x)-\pa_\nu \CA_\mu(t,x).
\eea
The momentum constraint  \bref{vp} is 
\bea
\vp^\nu(t,x)&=&{\it\Pi}^\nu(t,x)+\int dy~\chi(x^0,-y^0)\;
\CF^{\mu\nu}(t,y)\;\pa^y_\mu\D(x-y)
\nn\\
 &=&
{\it\Pi}^\nu(t,x)+\D(x^0)\CF^{0\nu}(t,x)\approx 0
\label{vpu1a}
\eea
and the constraint \bref{EOM} is 
\bea
\7\vp^\nu(t,x)&=&\pa_\mu \CF^{\mu\nu}(t,x)\approx 0.
\label{EOMu1a}
\eea
The generator of the {$U(1)$} transformation 
is given, using \bref{G}, by
\bea
G[\Lam]&=&\int dx[\;{\it\Pi}^\mu \pa_\mu \Lam\;].
\label{defGa}
\eea
It is expressed as a linear combination of the constraints, 
\bea
G[\Lam]&=&\int dx~\Lam\left[-(\pa_\mu\vp^\mu)-\D(x^0)\7\vp^0\right].
\label{GLama}
\eea
The Hamiltonian is expressed using the constraints and the $U(1)$ generator
as 
\bea
H&=&G[\CA_0]\;+\;\int dx\;\vp^i\;\CF_{0i}\;+\;
\int dx\; \D(x^0)\;\{\frac12 {\CF_{0i}}^2+\frac14 {\CF_{ij}}^2 \}.
\label{Hamb}
\eea
\medskip

The  Hamiltonian \bref{Hamb} as well as the 
constraints \bref{vpu1a} and \bref{EOMu1a} contain 
no time ($t$) derivative and are functions of the canonical pairs 
$(\CA_\mu(t,x),{\it\Pi}^\mu(t,x))$.
They are conserved since the Maxwell \lag in $d$ dimensions
has time translation invariance. The $U(1)$ generator is also conserved,
 without using constraints, for 
$\Lam(t,x)$ satisfying \bref{Lamdd}, 
\bea
\frac{d}{dt}G[\Lam]&=&\{G[\Lam],H\}~+~\frac{\pa}{\pa t}G[\Lam]~=~0
,~~~~~~~~~~~~~~
\dot \Lam=\pa_{x^0}\Lam.
\label{dtGLam2}
\eea
in agreement with \bref{dtG}.
Since the parameter $\Lam$ is subject to the last relation in \bref{dtGLam2}
the $U(1)$ transformations in the $\dpo$ dimensional canonical formulation 
are not {gauge} but {rigid} ones. We will see how the gauge 
transformations appear when it is written in a $d$ dimensional form.
\vskip 4mm

In cases where our  Lagrangians are local 
or higher derivative ones it is often
convenient to make expansion of the canonical variables 
using the Taylor basis\cite{Marnelius} in reducing them 
to $d$ dimensional forms. We expand the canonical variables as
\bea
\CA_\mu(t,x) &\equiv&\sum_{m=0}^{\infty}~e_m(x^0)~A_\mu^{(m)}(t,\bx),\;~~~~
\iPi^\mu(t,x) \equiv\sum_{m=0}^{\infty}~e^m(x^0)~\iPi^\mu_{(m)}(t,\bx),
\label{Taylor}\eea
where $e^\l(x^0)$ and $e_\l(x^0)$ are orthonormal basis 
\bea
e^\l(x^0)~=~(-\pa_{x^0})^\l\D(x^0),&&~~~~e_\l(x^0)~=~\frac{{(x^0)}^\l}{\l!},
\\
\int d{x^0}\; e^\l({x^0})~e_m({x^0})~=~\D^\l_m,&&~~~~
\sum_{\l=0}^{\infty}\; e^\l({x^0})~e_\l({x^0}')=\D({x^0}-{x^0}').
\eea
The $(A_\mu^{(m)}(t,\bx),\iPi^\mu_{(m)}(t,\bx))$ are $d$ dimensional
fields and are the new symplectic coordinates 
\bea
\Omega(t)&=&\int dx\; \delta \iPi^\mu(t,x)\wedge\delta \CA_\mu(t,x)~=~
\sum_{m=0}^{\infty}~\int d\bx~\delta \iPi^\mu_{(m)}(t,\bx)\wedge\D
A_\mu^{(m)}(t,\bx).
\nn\\
\label{symplectich}
\eea

In terms of them the constraint 
\bref{vpu1a} is expressed as
\bea
\vp^\mu(t,x)&=&\sum_{m=0}^{\infty}e^m(x^0)\vp^\mu_{(m)}(t,\bx), 
\\
\vp^0_{(m)}(t,\bx)&=&\iPi^0_{(m)}(t,\bx)~=~0
,~~~~~~~~~~~~~~~~~~~~~~~~~(m\geq 0),
\label{vp0m}\\
\vp^i_{(0)}(t,\bx)&=&\iPi^i_{(0)}(t,\bx)-
(\CA_i^{(1)}(t,\bx)-\pa_i\CA_0^{(0)}(t,\bx))~=~0,
\label{vpi0}\\
\vp^i_{(m)}(t,\bx)&=&\iPi^i_{(m)}(t,\bx)~=~0
,~~~~~~~~~~~~~~~~~~~~~~~~~(m\geq 1).
\label{vpim}
\label{vpu1acomp}
\eea
while the constraint \bref{EOMu1a} is 
\bea
\7\vp^\mu(t,x)&=&\sum_{m=0}^{\infty}e_m(x^0)\7\vp^{\mu{(m)}}(t,\bx),
\\
\7\vp^{i(m)}(t,\bx)&=&\pa_j(\pa_j\CA_i^{(m)}(t,\bx)-\pa_i\CA_j^{(m)}
(t,\bx))-(\CA_i^{(m+2)}(t,\bx)-\pa_i\CA_0^{(m+1)}(t,\bx))~=~0,
\label{tvpim}\nn\\
&&\hskip 70mm~~~~~~~~~~(m\geq 0),
\\
\7\vp^{0(m)}(t,\bx)&=&\pa_i(\CA_i^{(m+1)}(t,\bx)-\pa_i\CA_0^{(m)}(t,\bx)
)~=~0,~~~~~~~~~~~~~(m\geq 0).
\label{tvp0m}\label{EOM1acomp}
\eea
It must be noted the identities
\bea
\7\vp^{0(m+1)}(t,\bx)&=&\pa_i\7\vp^{i(m)}(t,\bx),~~~~~~~~~~~~~(m\geq 0).
\eea
Thus the only independent constraint of \bref{tvp0m} is $m=0$ 
case. It can be expressed, using \bref{vpi0}, as the gauss law constraint,
\bea
\7\vp^{0(0)}(t,\bx)&=&\pa_i
\iPi^i_{(0)}(t,\bx)~=~0.~
\label{gauss}\eea

Following the Dirac's standard procedure of constraints \cite{Dirac} 
we classify them
and eliminate the second class constraints.
The constraints \bref{vpim} $(m\geq 2)$ are paired with 
the constraints \bref{tvpim} $(m\geq 0)$ to form second class sets.
They are used to eliminate canonical pairs 
$(\CA_i^{(m)}(t,\bx),\iPi^i_{(m)}(t,\bx)),(m\geq 2)$ as
\bea
\CA_i^{(m)}(t,\bx)&=&
\pa_j(\pa_j\CA_i^{(m-2)}(t,\bx)-\pa_i\CA_j^{(m-2)}
(t,\bx))+\pa_i\CA_0^{(m-1)}(t,\bx),
\\
\iPi^i_{(m)}(t,\bx))&=&0,~~~~~~~~~~~(m\geq 2).
\eea 
The constraints \bref{vpim} $(m=1)$ and
\bref{vpi0} are paired to a second class set and
are used to eliminate 
$(\CA_i^{(1)}(t,\bx),\iPi^i_{(1)}(t,\bx))$ as
\bea
\CA_i^{(1)}(t,\bx)&=&\iPi^i_{(0)}(t,\bx)+\pa_i\CA_0^{(0)}(t,\bx),~
\\
\iPi^i_{(1)}(t,\bx)&=&0.
\eea 

After eliminating the canonical pairs 
$(\CA_i^{(m)}(t,\bx),\iPi^i_{(m)}(t,\bx)),(m\geq 1)$ 
using the second class constraints the system is described
in terms of the canonical pairs 
$(\CA_i^{(0)}(t,\bx),\iPi^i_{(0)}(t,\bx))$ and 
$(\CA_0^{(m)}(t,\bx),\iPi^0_{(m)}(t,\bx)),(m\geq 0)$.
The Dirac brackets among them remain same as the Poisson brackets.
Remember the $d$ dimensional fields are identified by \bref{qdef} as
\bea
A_\mu(t,\bx)&=&\CA_\mu(t,0,\bx)~=~\CA_\mu^{(0)}(t,\bx),~~~~~~~
\Pi^\mu(t,\bx)~=~\iPi^\mu_{(0)}(t,\bx).
\eea
The remaining constraints are \bref{gauss} and \bref{vp0m},
\bea
\pa_i\iPi^i_{(0)}(t,\bx)~=~0,~~~~~~~~~
\iPi^0_{(m)}(t,\bx)~=~0.~~~~~~~~~(m\geq 0)
\label{firstcc}
\eea
They are first class constraints. 
The Hamiltonian \bref{Hamb} in the reduced variables is
\bea
H(t)&=&\int d\bx\left[~
\sum_{m=0}^{\infty}~\CA_0^{(m+1)}(t,\bx)\iPi^0_{(m)}(t,\bx)~-~
\CA_0^{(0)}(t,\bx)(\pa_i\iPi^i_{(0)}(t,\bx))\right.
\nn\\
&&\left.~+~
\frac12 (\iPi^i_{(0)}(t,\bx))^2~+~
\frac14 (\pa_j\CA_i^{(0)}(t,\bx)-\pa_i\CA_j^{(0)}(t,\bx))^2\right].
\label{hh}
\eea
The $U(1)$ generator \bref{GLama} is
\bea
G[\Lam]&=&\int d\bx~
\left[~
\sum_{m=0}^{\infty}~\Lam^{(m+1)}(t,\bx)\iPi^0_{(m)}(t,\bx)~-~
\Lam^{(0)}(t,\bx)(\pa_i\iPi^i_{(0)}(t,\bx))\right],
\label{Lamlam}
\eea
where
\bea
\Lam(t,\lam)&=& \sum_{m=0}^{\infty}~\Lam^{(m)}(t,\bx)e_m(x^0),~~~~~~~
{\rm and}~~~~~~~\dot\Lam^{(m)}(t,\bx)=\Lam^{(m+1)}(t,\bx).
\eea

The first class constraints $\iPi^0_{(m)}(t,\bx)=0,(m\geq 0)$ in 
\bref{firstcc} mean that $\CA_0^{(m)}(t,\bx),$ $(m\geq 0)$ are the
gauge degrees of freedom and we can  assign to them 
{ any function of $\bx$
for all values of $m$ at given time} 
$t=t_0$. It is equivalent to saying that we can assign 
{any function of time
to $\CA_0^{(0)}(t,\bx)$ for all value of $t$}, 
due to the equation of motion
$\dot\CA_0^{(m)}(t,\bx)=\CA_0^{(m+1)}(t,\bx)$.
In this way we can understand that the Hamiltonian \bref{hh}
is equivalent to the standard form of the canonical Hamiltonian of 
the Maxwell theory,
\bea
H(t)&=&\int d\bx\left[~
\dot A_0(t,\bx)\Pi^0(t,\bx)~-~
A_0(t,\bx)(\pa_i\Pi^i(t,\bx))
\right.\nn\\&&\left.~+~\frac12 (\Pi^i(t,\bx))^2~+~
\frac14 (\pa_jA_i(t,\bx)-\pa_iA_j(t,\bx))^2\right]
\label{hhm}
\eea
in which $A_0(t,\bx)$ is arbitrary function of time.
In the same manner the $U(1)$ generator \bref{Lamlam} is
\bea
G[\Lam]&=&\int d\bx~
\left[~
\dot\lam(t,\bx)\Pi^0(t,\bx)~-~
\lam(t,\bx)(\pa_i\Pi^i(t,\bx))\right],
\label{Lamlamm}
\eea
in which the gauge parameter function $\lam(t,\bx)\equiv\Lam^{(0)}(t,\bx)$ 
is regarded as any function of time.
\vskip 6mm



\begin{thebibliography}{99}
\bibitem{lv} J. Llosa and J. Vives, {\it J. Math. Phys.} 
{\bf 35} (1994) 2856.   
\bibitem{gkl} J.  Gomis, K.  Kamimura and J.  Llosa, {\it Phys. Rev.D} to 
appear, {\bf [hep-th/0006235]}. 
\bibitem{woodd1} R.  P.  Woodard, {\bf [hep-th/0006207]}. 
\bibitem{bearing} K.  Bering, {\bf [hep-th/0007192]}. 
\bibitem{sw} N. Seiberg and E. Witten, 
{\it J. High. Energy Phys.}  {\bf 9909} (1999) 032, {\bf [hep-th/9908142]}. 
\bibitem{Dirac}P. A. M.  Dirac, {\it Can.  J.  Math.}  {\bf 2} (1950) 129. 
\bibitem{o} M.  Ostrogradski, {\it Mem.  Ac.  St.  Peterbourg} 
{\bf 6} (1850) 385. 
\bibitem{nother} E.  Noether, {\it Nachr.  Ges.  Wiss.  Gettinger,} {\bf 2} 
(1918) 235. 
\bibitem{kp} N. P. Konopleva and V. N. Popov, {\it Gauge Fields}
 (Harwood Academic Pub. ) (1981). 
\bibitem{GKP}
J.  Gomis, K.  Kamimura and J.  Pons, {\it Europhysics Lett.}  
{\bf  2} (1986) 187. 
\bibitem{josepmaria} J.  M.  Pons and J.  Antonio Garc\'{\i}a, 
{\it Int. J. Mod. Phys.} {\bf A15} (2000) 4681, {\bf [hep-th/9908151]}. 
\bibitem{gkll2} J.  Gomis, K.  Kamimura and J.  Llosa in preparation. 
\bibitem{sst} N. Seiberg, L. Susskind and N. Toumbas, {\it 
J. High. Energy Phys.  }{\bf 0006} (2000) 044,
{\bf [hep-th/0005015]}. 
\bibitem{ab} L. Alvarez-Gaum\'e and J. L. F. Barb\'on, {\bf [hep-th/0006209]}. 
\bibitem{gm} J. Gomis and T. Mehen, {\it Nucl. Phys.} {\bf B591} (2000) 265,  
{\bf [hep-th/0005129]}. 
\bibitem{agm} O. Aharony, J. Gomis and T. Mehen, {\it 
J. High. Energy Phys.  }{\bf 0009} (2000) 023, {\bf [hep-th/0006236]}. 
\bibitem{sst1} N. Seiberg, L. Susskind and N.  Toumbas. 
{\it J. High.  Energy Phys.}  {\bf 0006 }  (2000) 021, {\bf [hep-th/0005040]}. 
\bibitem{gms} R. Gopakumar, J. Maldacena, S. Minwalla and A. Strominger,
{\it J. High. Energy Phys.}  {\bf 0006} (2000) 036, {\bf [hep-th/0054048]}. 
\bibitem{br} J. L. F. Barbon and E. Rabinovici, {\it Phys. Lett.}  
{\bf B486} (2000) 202,{\bf [hep-th/0005073]}. 
\bibitem{gmms}  R. Gopakumar, S. Minwalla, N. Seiberg and A. Strominger,
{\it J. High. Energy Phys.}  {\bf 0008} (2000) 008,
{\bf [hep-th/0006062]}. 
\bibitem{km} I. R. Klebanov and J. Maldacena, {\bf [hep-th/0006085]}. 
\bibitem{Dayi} O. F. Dayi, {\it Phys. Lett.} {\bf B481} (2000) 408,
{\bf [hep-th/0001218]}.
\bibitem{brs} C.  Becchi, A.  Rouet and R.  Stora, {\it Phys.  Lett. } {\bf
52B} (1974) 344. 
\bibitem{t} I.  V.  Tyutin, Lebedev preprint (1975). 
\bibitem{z} J.  Zinn-Justin, in {\it Trends in Elementary Particle Theory,}
International Summer Institute on Theoretical Physics in Bonn 1974. 
Springer-Verlag 1975
\bibitem{bv} I.  A.  Batalin and G.  A.  Vilkovisky, {\it Phys.  Lett. } {\bf
102B} (1981) 27. 
\bibitem{ht} M.  Henneaux and C.  Teitelboim, {\it Quantization of Gauge
Systems}, Princeton University Press, Princeton (1992). 
\bibitem{gps} J.  Gomis, J.  Par\'{i}s and S. Samuel, {\it Phys. Rept.}  {\bf
259} (1995) 1. 
\bibitem{w} S. Weinberg , {\it The Quantum Theory of Fields} , vol.2,  
Cambridge University Press, Princeton (1996).
\bibitem{ku} T.  Kugo and S.  Uehara, {\it Nucl.  Phys. } 
{\bf B197} (1982) 378. 
\bibitem{bv85}I.  A.  Batalin and G.  A.  Vilkovisky, 
{\it J. Math. Phys.} {\bf 26} (1985) 172. 
\bibitem{fh90} J.  M.  L Fisch and M.  Henneaux.  {\it Comm.  Math.  Phys.} 
{\bf 128}(1990) 627. 
\bibitem{fv75} E.  S.  Fradkin and G.  A.  Vilkovisky, {\it Phys.  Lett.}  {\bf
55B} (1975) 224. 
\bibitem{bv77} I.  A.  Batalin and G.  A.  Vilkovisky, {\it Phys.  Lett.}  {\bf
69B} (1977) 309. 
\bibitem{roberto} R.  Casalbuoni, {\it Nuovo Cim.}  {\bf A33} (1976) 389.  
\bibitem{vt} B. L.  Voronov and I.  V.  Tyutin, {\it Theor.  Math.  Phys.}
  {\bf 50} (1982) 218. 
\bibitem{Marnelius} R. Marnelius, {\it Phys.  Rev.}  {\bf D10} (1974) 2535.  


\end{thebibliography}
\end{document}